\documentclass[12pt]{article}
\usepackage{a4wide}
\usepackage{latexsym}
\usepackage{amsmath}
\usepackage{amsfonts}
\usepackage{amscd}
\usepackage{cite}

\usepackage{pslatex}
\usepackage{graphicx}
\usepackage[latin1]{inputenc}
\usepackage[T1]{fontenc}

\allowdisplaybreaks

\newcommand{\bq}{\begin{eqnarray}}
\newcommand{\eq}{\end{eqnarray}}
\newcommand{\eps}{\varepsilon}

\begin{document}

\thispagestyle{empty}

\begin{flushright}
  MITP/15-114 
 \\ MaPhy-AvH/2015-21
\end{flushright}

\vspace{1.5cm}

\begin{center}
  {\Large\bf The iterated structure of the all-order result for the two-loop sunrise integral \\
  }
  \vspace{1cm}
  {\large Luise Adams ${}^{a}$, Christian Bogner ${}^{b}$ and Stefan Weinzierl ${}^{a}$ \\
  \vspace{1cm}
      {\small ${}^{a}$ \em PRISMA Cluster of Excellence, Institut f{\"u}r Physik, }\\
      {\small \em Johannes Gutenberg-Universit{\"a}t Mainz,}\\
      {\small \em D - 55099 Mainz, Germany}\\
  \vspace{2mm}
      {\small ${}^{b}$ \em Institut f{\"u}r Physik, Humboldt-Universit{\"a}t zu Berlin,}\\
      {\small \em D - 10099 Berlin, Germany}\\
  } 
\end{center}

\vspace{2cm}

\begin{abstract}\noindent
  {
We present a method to compute the Laurent expansion of the two-loop sunrise integral
with equal non-zero masses to arbitrary order in the dimensional regularisation $\eps$.
This is done by introducing a class of functions (generalisations of multiple polylogarithms to include the elliptic case)
and by showing that all integrations can be carried out within this class of functions.
   }
\end{abstract}

\vspace*{\fill}

\newpage

\section{Introduction}
\label{sec:intro}

Feynman integrals are uttermost important for precision calculations in particle physics.
Due to the presence of ultraviolet or infrared divergences these integrals may require regularisation.
It is common practice to use dimensional regularisation \cite{'tHooft:1972fi,Bollini:1972ui,Cicuta:1972jf}
with regularisation parameter $\eps$ and to
present the result for a Feynman integral as a Laurent series in $\eps$.
It is a natural question to ask, what transcendental functions appear in the $\eps^j$-term.
For one-loop integrals and for the expansion around four space-time dimensions the answer for the $\eps^0$-term
is simple: There are just two transcendental functions. These are the logarithm
\bq
 \mathrm{Li}_1\left(x\right) & = & - \ln\left(1-x\right)
 \;\; = \;\;
 \sum\limits_{n=1}^\infty \frac{x^n}{n},
\eq
and the dilogarithm
\bq
 \mathrm{Li}_2\left(x\right) & = & 
 \sum\limits_{n=1}^\infty \frac{x^n}{n^2}.
\eq
Of course we would like to generalise this to multi-loop integrals, to expansions around any even space-time
dimension and to arbitrary order $\eps^j$.
There is a wide class of Feynman integrals for which this can be done.
These Feynman integrals evaluate to generalisations of the two transcendental functions above,
called multiple polylogarithms.
The multiple polylogarithms are defined by \cite{Goncharov_no_note,Goncharov:2001,Borwein}
\bq
 \mathrm{Li}_{n_1,n_2,...,n_k}\left(x_1,x_2,...,x_k\right)
 & = &
 \sum\limits_{j_1=1}^\infty \sum\limits_{j_2=1}^{j_1-1} ... \sum\limits_{j_k=1}^{j_{k-1}-1}
 \frac{x_1^{j_1}}{j_1^{n_1}} \frac{x_2^{j_2}}{j_2^{n_2}} ... \frac{x_k^{j_k}}{j_k^{n_k}}.
\eq
The multiple polylogarithms have also a representation as iterated integrals
and enjoy several nice algebraic properties: There is a shuffle and a quasi-shuffle product, derived from 
the integral and sum representation, respectively. Methods for the numerical evaluation are available \cite{Vollinga:2004sn}.
This allows that a wide class of Feynman integrals can be computed 
systematically to all orders in $\eps$.
Algorithms which accomplish this are for example based on nested sums \cite{Moch:2001zr,Weinzierl:2002hv,Weinzierl:2004bn,Moch:2005uc,Bierenbaum:2003ud},
linear reducibility \cite{Brown:2008,Panzer:2014caa,Bogner:2015unknown}
or differential equations \cite{Kotikov:1990kg,Kotikov:1991pm,Remiddi:1997ny,Gehrmann:1999as,Argeri:2007up,MullerStach:2012mp,Henn:2013pwa}.
On the mathematical side, multiple polylogarithms are closely related to punctured Riemann surfaces of genus zero \cite{Goncharov:2001,Brown:2006,Bogner:2014mha}.

There are however Feynman integrals, which cannot be expressed in multiple polylogarithms.
The aim of this paper is to study how the class of multiple polylogarithms needs to be extended.
The simplest Feynman integral which cannot be expressed in terms of multiple polylogarithms is the two-loop
sunrise integral with non-vanishing masses.
This Feynman integral has already received considerable attention 
in the literature \cite{Broadhurst:1993mw,Berends:1993ee,Bauberger:1994nk,Bauberger:1994by,Bauberger:1994hx,Caffo:1998du,Laporta:2004rb,Groote:2005ay,Groote:2012pa,Bailey:2008ib,MullerStach:2011ru,Adams:2013nia,Bloch:2013tra,Remiddi:2013joa,Adams:2014vja,Adams:2015gva,Caffo:2002ch,Pozzorini:2005ff,Caffo:2008aw}.
In this paper we study the two-loop sunrise integral with equal non-zero masses in $D=2-2\eps$ space-time dimensions.
Considering the sunrise integral in $2-2\eps$ dimensions instead of $4-2\eps$ dimensions is no restriction:
With the help of dimensional recurrence relations \cite{Tarasov:1996br,Tarasov:1997kx} one recovers the result in $4-2\eps$ dimensions
from the result in $2-2\eps$ dimensions.
The explicit equation for the dimensional shift can be found in eq.~(114) of \cite{Adams:2015gva}.
Working in $D=2-2\eps$ dimensions has the advantage that the leading term in the $\eps$-expansion (in $D=2-2\eps$ dimensions the two-loop sunrise integral
is finite and the $\eps$-expansions starts with $\eps^0$) depends only on one graph polynomial, but not both.
The zero locus of this graph polynomial describes an elliptic curve, i.e. a Riemann surface of genus one.
The higher terms in the $\eps$-expansion depend of course on both graph polynomials. The other graph polynomial describes a Riemann surface of genus zero.
This gives us already an indication, what transcendental functions we should expect: As multiple polylogarithms correspond to the pure genus zero case,
we are looking for a generalisation which corresponds to a combination of genus one and genus zero.
The lowest term in the $\eps$-expansion will correspond to a pure genus one case.

In this paper we define a class of functions, which can be seen as a generalisation of the multiple polylogarithms to the mixed genus one / genus zero case.
This class includes the multiple polylogarithms.
The new additional functions are
\bq
\lefteqn{
 \mathrm{ELi}_{n_1,...,n_l;m_1,...,m_l;2o_1,...,2o_{l-1}}\left(x_1,...,x_l;y_1,...,y_l;q\right) 
 = }
 & & \nonumber \\
 & = &
 \sum\limits_{j_1=1}^\infty ... \sum\limits_{j_l=1}^\infty
 \sum\limits_{k_1=1}^\infty ... \sum\limits_{k_l=1}^\infty
 \;\;
 \frac{x_1^{j_1}}{j_1^{n_1}} ... \frac{x_l^{j_l}}{j_l^{n_l}}
 \;\;
 \frac{y_1^{k_1}}{k_1^{m_1}} ... \frac{y_l^{k_l}}{k_l^{m_l}}
 \;\;
 \frac{q^{j_1 k_1 + ... + j_l k_l}}{\prod\limits_{i=1}^{l-1} \left(j_i k_i + ... + j_l k_l \right)^{o_i}}.
\eq
We show that each term of the $\eps$-expansion of the two-loop sunrise integral can be expressed in this class of functions and we give an algorithm
to compute the term of order $\eps^j$.
In an appendix we present the explicit results for the first three terms of the $\eps$-expansion.

On a technical level this is achieved by considering the differential equation for the two-loop sunrise integral.
We bring the differential equation to a particularly useful form, such that all integrations (after an appropriate change of variables)
can be carried out easily.
This step is inspired by Henn's method \cite{Henn:2013pwa} in the genus zero case.
In this particular form of the differential equation it is straightforward to show that we always stay within the specific class of functions.
We would like to stress that the change of variables from the momentum squared $t=p^2$ to the nome of the elliptic curve $q$ is crucial \cite{Bloch:2013tra,Adams:2014vja}:
We will see that all integrands have nice forms in the variable $q$.

This paper is organised as follows:
In section~\ref{sec:notation} we define the two-loop sunrise integral and we introduce our notation.
In section~\ref{sec:elliptic_variables} we review variables naturally associated to the elliptic curve.
In section~\ref{sec:dgl_trafo} we derive a particularly useful form of the differential equation for the
two-loop sunrise integral with equal masses around two space-time dimensions.
This form allows us to obtain the $\eps^j$-term of the Laurent expansion from lower order terms and 
(simple) integrations.
The solution of a differential equation requires in addition boundary values.
These are discussed in section~\ref{sec:boundary_values}.
In section~\ref{sec:functions} we introduce the class of functions, in which the two-loop sunrise integral can be expressed.
Section~\ref{sec:algorithm} contains the main result of this paper and gives an algorithm 
to compute the $\eps^j$-term in the Laurent expansion of the two-loop sunrise integral.
Finally, our conclusions are given in section~\ref{sec:conclusions}.
In an appendix we present the explicit results for the first three terms of the $\eps$-expansion of the two-loop
sunrise integral.

\section{Basic set-up}
\label{sec:notation}

The two-loop integral corresponding to the sunrise graph with equal masses is given 
in $D$-dimensional Minkowski space by
\bq
\label{def_sunrise}
\lefteqn{
 S_{111}\left( D, p^2, m^2, \mu^2 \right)
 = 
} & &
 \\
 & &
 \left(\mu^2\right)^{3-D}
 \int \frac{d^Dk_1}{i \pi^{\frac{D}{2}}} \frac{d^Dk_2}{i \pi^{\frac{D}{2}}}
 \frac{1}{\left(-k_1^2+m^2\right) \left(-k_2^2+m^2\right) \left(-\left(p-k_1-k_2\right)^2+m^2\right)}.
 \nonumber
\eq
The arbitrary scale $\mu$ is introduced to keep the integral dimensionless.
The quantity $p^2$ denotes the momentum squared (with respect to the Minkowski metric)
and we will write
\bq
 t & = & p^2.
\eq
Where it is not essential we will suppress the dependence on the mass $m$ and the scale $\mu$ and simply write
$S_{111}( D, t)$ instead of $S_{111}( D, t, m^2, \mu^2)$.
In terms of Feynman parameters the two-loop integral is given by
\bq
\label{def_Feynman_integral}
 S_{111}\left( D, t\right)
 & = & 
 \Gamma\left(3-D\right)
 \left(\mu^2\right)^{3-D}
 \int\limits_{\sigma} 
 \frac{{\cal U}^{3-\frac{3}{2}D}}{{\cal F}^{3-D}} \omega
\eq
with the two Feynman graph polynomials
\bq
 {\cal U} \; = \; x_1 x_2 + x_2 x_3 + x_3 x_1,
 & &
 {\cal F} \; = \; - x_1 x_2 x_3 t + m^2 \left( x_1 + x_2 + x_3 \right) {\cal U}.
\eq
The differential two-form $\omega$ is given by
\bq
 \omega & = & x_1 dx_2 \wedge dx_3 + x_2 dx_3 \wedge dx_1 + x_3 dx_1 \wedge dx_2,
\eq
and the integration is over
\bq
 \sigma & = & \left\{ \left[ x_1 : x_2 : x_3 \right] \in {\mathbb P}^2 | x_i \ge 0, i=1,2,3 \right\}.
\eq
For the sunrise integral any integral obtained by pinching propagators is a product of tadpole integrals.
The tadpole integral is given by
\bq
\label{def_tadpole}
 T_{1}\left( D, m^2, \mu^2 \right)
 & = &
 \left(\mu^2\right)^{1-\frac{D}{2}}
 \int \frac{d^Dk}{i \pi^{\frac{D}{2}}}
 \frac{1}{\left(-k^2+m^2\right)}
 \;\; = \;\;
 \Gamma\left(1-\frac{D}{2}\right)
 \left( \frac{m^2}{\mu^2} \right)^{\frac{D}{2}-1}.
\eq
It has been known for a long time that in the equal mass case 
the two-loop sunrise integral satisfies a second-order differential equation for all values of $D$ \cite{Broadhurst:1993mw,Laporta:2004rb}:
\bq
\label{dgl_D_dim}
 \left( p_2 \frac{d^2}{dt^2} + p_1 \frac{d}{dt} + p_0 \right)
 S_{111}\left(D,t\right)
 = 
 - 6 \frac{m^4}{\mu^2} \left[ T_{1}\left(D-2\right) \right]^2,
\eq
with $T_{1}(D-2)=T_{1}(D-2,m^2,\mu^2)$ and
\bq
 p_2 & = & t \left(t-9m^2\right) \left(t-m^2\right),
 \nonumber \\
 p_1 & = & \frac{3}{2} \left(4-D\right) t^2 + 5 \left(D-6\right) t m^2 + \frac{9}{2} D m^4,
 \nonumber \\
 p_0 & = & \frac{\left(D-3\right)}{2} \left[ \left(D-4\right) t + \left(D+4\right) m^2 \right].
\eq

\section{Variables related to the elliptic curve}
\label{sec:elliptic_variables}

We may view the graph polynomial ${\mathcal F}$ as a polynomial in the Feynman parameters 
$x_1$, $x_2$, $x_3$ with parameters $t$ and $m^2$.
The algebraic equation
\bq
 {\mathcal F} & = & 0
\eq
defines together with the choice of a rational point as origin an elliptic curve.
The modulus $k$ and the complementary modulus $k'$ of the elliptic curve are given by
\bq
\label{def_modulus}
 k = \sqrt{\frac{e_3-e_2}{e_1-e_2}},
 & &
 k' = \sqrt{1-k^2} = \sqrt{\frac{e_1-e_3}{e_1-e_2}}.
\eq
The variables $e_1$, $e_2$ and $e_3$ are the roots of the cubic polynomial of the Weierstrass normal form
$y^2=4(x-e_1)(x-e_2)(x-e_3)$ and given in terms of the parameters of the elliptic curve by
\bq
\label{def_roots}
 e_1 
 & = & 
 \frac{1}{24 \mu^4} \left( -t^2 + 6 m^2 t + 3 m^4 + 3 \sqrt{\tilde{D}} \right),
 \nonumber \\
 e_2 
 & = & 
 \frac{1}{24 \mu^4} \left( -t^2 + 6 m^2 t + 3 m^4 - 3 \sqrt{\tilde{D}} \right),
 \nonumber \\
 e_3 
 & = & 
 \frac{1}{24 \mu^4} \left( 2 t^2 - 12 m^2 t - 6 m^4 \right).
\eq
As abbreviation we used
\bq
\label{def_D}
 \tilde{D} & = & 
 \left( t - m^2 \right)^3
 \left( t - 9 m^2 \right).
\eq
The periods of the elliptic curve can be taken as
\bq
\label{def_periods}
 \psi_1 =  
 2 \int\limits_{e_2}^{e_3} \frac{dx}{y}
 =
 \frac{4 \mu^2}{\tilde{D}^{\frac{1}{4}}} K\left(k\right),
 & &
 \psi_2 =  
 2 \int\limits_{e_1}^{e_3} \frac{dx}{y}
 =
 \frac{4 i \mu^2}{\tilde{D}^{\frac{1}{4}}} K\left(k'\right),
 \\
 \phi_1 =  
 \frac{8\mu^4}{\tilde{D}^{\frac{1}{2}}} \int\limits_{e_2}^{e_3} \frac{\left(x-e_2\right) dx}{y}
 =
 \frac{4 \mu^2}{\tilde{D}^{\frac{1}{4}}} \left( K\left(k\right)- E\left(k\right) \right),
 & &
 \phi_2 =  
 \frac{8\mu^4}{\tilde{D}^{\frac{1}{2}}} \int\limits_{e_1}^{e_3} \frac{\left(x-e_2\right) dx}{y}
 =
 \frac{4 i \mu^2}{\tilde{D}^{\frac{1}{4}}} E\left(k'\right).
 \nonumber
\eq
$K(x)$ and $E(x)$ denote the complete elliptic integral of the first kind and second kind, respectively:
\bq
 K(x)
 = 
 \int\limits_0^1 \frac{dt}{\sqrt{\left(1-t^2\right)\left(1-x^2t^2\right)}},
 & &
 E(x)
 = 
 \int\limits_0^1 dt \sqrt{\frac{1-x^2t^2}{1-t^2}}.
\eq
We denote the ratio of the two periods $\psi_2$ and $\psi_1$ by
\bq
 \tau 
 & = & 
 \frac{\psi_2}{\psi_1}
\eq
and the nome by
\bq
\label{def_nome}
 q & = & e^{i\pi \tau}.
\eq
We may express the variable $t$ as a function of the nome $q$ 
(and we will use this change of variables extensively in the rest of the paper):
\bq
 t & = & 
 - 9 m^2 
 \frac{\eta\left(\tau\right)^4 \eta\left(\frac{3\tau}{2}\right)^4 \eta\left(6\tau\right)^4}
      {\eta\left(\frac{\tau}{2}\right)^4 \eta\left(2\tau\right)^4 \eta\left(3\tau\right)^4},
\eq
where $\eta(\tau)$ denotes Dedekind's $\eta$-function
\bq
 \eta\left(\tau\right)
 & = &
 e^{\frac{\pi i \tau}{12}} \prod\limits_{n=1}^\infty \left( 1- e^{2 \pi i n \tau} \right)
 =
 q^{\frac{1}{12}} \prod\limits_{n=1}^\infty \left( 1 - q^{2n} \right).
\eq
The first few terms read
\bq
 t 
 & = & 
 - 9 m^2 q
 - 36 m^2 q^2
 - 90 m^2 q^3
 - 180 m^2 q^4
 + {\mathcal O}\left(q^5\right).
\eq
The Wronskian is given by
\bq
\label{def_Wronski}
 W & = &
 \psi_1 \frac{d}{dt} \psi_2 - \psi_2 \frac{d}{dt} \psi_1
 =
 -
 \frac{12 \pi i \mu^4}{t\left( t - m^2 \right)\left( t - 9 m^2 \right)}.
\eq
We further denote by $r_n$ the $n$-th root of unity
\bq
 r_n & = & e^{\frac{2\pi i}{n}},
\eq
In particular we will need the third root of unity
\bq
 r_3 & = & e^{\frac{2 \pi i}{3}} 
 \;\; = \;\;
 \frac{1+i\sqrt{3}}{1-i\sqrt{3}}
 \;\; = \;\;
 - \frac{1}{2} + \frac{i}{2} \sqrt{3}.
\eq

\section{A particularly useful form of the differential equation}
\label{sec:dgl_trafo}

We study the two-loop sunrise integral around two space-time dimensions.
We therefore set $D=2-2\eps$.
The Feynman parameter integral reads
\bq
 S_{111}\left( 2-2\eps, t\right)
 & = & 
 \Gamma\left(1+2\eps\right)
 \left(\mu^2\right)^{1+2\eps}
 \int\limits_{\sigma} 
 \frac{{\cal U}^{3\eps}}{{\cal F}^{1+2\eps}} \omega.
\eq
The algebraic curve defined by ${\mathcal F}=0$ has genus $1$, the algebraic curve
defined by ${\mathcal U}=0$ has genus $0$.
The algebraic curve given by ${\mathcal F}=0$ defines 
together with the choice of a rational point as origin an elliptic curve.
At order $\eps^0$ the two-loop sunrise integral 
depends only on the graph polynomial ${\mathcal F}$, but not on the graph
polynomial ${\mathcal U}$.
This suggests (and is justified a posteriori) that the variables related to the elliptic curve introduced
in the previous section are convenient variables to express the result of the two-loop sunrise integral.
At order $\eps^1$ and beyond the two-loop sunrise integral depends on both graph polynomials 
${\mathcal F}$ and ${\mathcal U}$.
Still, it is convenient to keep the variables from the $\eps^0$-case.

The sunrise integral has a Taylor expansion in $\eps$:
\bq
\label{expansion_2D}
  S_{111}\left( 2-2\eps, t\right)
 & = &
 e^{-2 \gamma \eps} \sum\limits_{j=0}^\infty \eps^j S_{111}^{(j)}(2,t).
\eq
The differential equation given in eq.~(\ref{dgl_D_dim}) reads in $D=2-2\eps$ dimensions
\bq
\label{dgl_2_2eps}
 L_2  S_{111}\left(2-2\eps,t\right)
 & = &
 - 6 \mu^2 \Gamma\left(1+\eps\right)^2 
 \left( \frac{\mu^2}{m^2} \right)^{2\eps}.
\eq
The Picard-Fuchs operator $L_2$ has the $\eps$-expansion
\bq
\label{expansion_L2_2D}
 L_2
 & = &
 \sum\limits_{j=0}^2 \eps^j L_2^{(j)},
\eq
with
\bq
\label{def_L2_expansion}
 L_2^{(0)}
 = 
 p_{2}^{(0)} \frac{d^2}{dt^2}
 +
 p_{1}^{(0)} \frac{d}{dt}
 +
 p_{0}^{(0)},
 \;\;\;\;\;\;
 L_2^{(1)}
 = 
 p_{1}^{(1)} \frac{d}{dt}
 +
 p_{0}^{(1)},
 \;\;\;\;\;\;
 L_2^{(2)}
 = 
 p_{0}^{(2)},
\eq
and
\begin{align}
 p_{2}^{(0)} & = t \left(t-m^2\right) \left(t-9m^2\right),
 &
 p_{1}^{(0)} & = 3 t^2 - 20 t m^2 + 9 m^4,
 &
 p_{0}^{(0)} & = t- 3 m^2,
 \\
 & &
 p_{1}^{(1)} & = 3 t^2 - 10 t m^2 - 9 m^4,
 &
 p_{0}^{(1)} & =  3t- 5 m^2,
 \nonumber \\
 & &
 & &
 p_{0}^{(2)} & = 2t+ 2 m^2.
 \nonumber 
\end{align}
We may view eq.~(\ref{dgl_2_2eps}) as a second-order differential equation for the $\eps^j$-term
$S_{111}^{(j)}(2,t)$, where in the inhomogeneous term 
in addition to the right-hand side of eq.~(\ref{dgl_2_2eps})
the lower order terms 
$S_{111}^{(j-1)}(2,t)$ and $S_{111}^{(j-2)}(2,t)$ together with the derivative $dS_{111}^{(j-1)}(2,t)/dt$ appear.
It is possible to simplify the differential equation by eliminating $L_2^{(1)}$. 
This implies the elimination of $S_{111}^{(j-1)}(2,t)$ and $dS_{111}^{(j-1)}(2,t)/dt$.
We set
\bq
\label{def_Stilde}
  S_{111}\left( 2-2\eps, t\right)
 & = &
 \Gamma\left(1+\eps\right)^2 
 \left( \frac{3 \mu^4 \sqrt{t}}{m \left(t-m^2\right) \left(t-9m^2\right)} \right)^{\eps}
 \tilde{S}_{111}\left( 2-2\eps, t\right).
\eq
The differential equation for $\tilde{S}_{111}(2-2\eps,t)$ reads then
\bq
\label{dgl_tilde_2_2eps}
 \tilde{L}_2  \tilde{S}_{111}\left(2-2\eps,t\right)
 & = &
 - 6 \mu^2 
 \left( \frac{\left(t-m^2\right) \left(t-9m^2\right)}{3 m^3 \sqrt{t}} \right)^{\eps}.
\eq
$\tilde{S}_{111}\left(2-2\eps,t\right)$ has a Taylor expansions in $\eps$,
which we write as
\bq
\label{expansion_tilde_2D}
  \tilde{S}_{111}\left( 2-2\eps, t\right)
 & = &
 \sum\limits_{j=0}^\infty \eps^j \tilde{S}_{111}^{(j)}(2,t).
\eq
In comparison to eq.~(\ref{expansion_2D}) we didn't factor out a prefactor $\exp(-2 \gamma \eps)$.
This is just a convenient convention. In both cases our definitions are such that the Taylor terms
$S_{111}^{(j)}(2,t)$ and $\tilde{S}_{111}^{(j)}(2,t)$ are free of Euler's constant $\gamma$.
The differential operator $\tilde{L}_2$ has a Taylor expansions in $\eps$ similar
to eq.~(\ref{expansion_L2_2D}) and we find
\begin{align}
 \tilde{p}_{2}^{(0)} & = t \left(t-m^2\right) \left(t-9m^2\right),
 &
 \tilde{p}_{1}^{(0)} & = 3 t^2 - 20 t m^2 + 9 m^4,
 &
 \tilde{p}_{0}^{(0)} & = t- 3 m^2,
 \\
 & &
 \tilde{p}_{1}^{(1)} & = 0,
 &
 \tilde{p}_{0}^{(1)} & = 0,
 \nonumber \\
 & &
 & &
 \tilde{p}_{0}^{(2)} & = -\frac{\left(t+3m^2\right)^4}{4 t \left(t-m^2\right)\left(t-9m^2\right)}.
 \nonumber 
\end{align}
Eq.~(\ref{dgl_tilde_2_2eps}) is a second-order differential equation for 
$\tilde{S}_{111}^{(j)}(2,t)$, where in the inhomogeneous term 
in addition to the right-hand side of eq.~(\ref{dgl_tilde_2_2eps})
only the  lower order term $\tilde{S}_{111}^{(j-2)}(2,t)$ appears.
The terms $\tilde{S}_{111}^{(j-1)}(2,t)$ and $d\tilde{S}_{111}^{(j-1)}(2,t)/dt$ do not appear.
We further note that the ${\mathcal O}(\eps^0)$-part of the Picard-Fuchs operators
$\tilde{L}_2$ and $L_2$ agree,
in other words we have $\tilde{L}_2^{(0)}=L_2^{(0)}$.
The differential equation for $\tilde{S}_{111}^{(j)}(2,t)$ is therefore
\bq
\label{dgl_transformed}
 L^{(0)}_{2} \tilde{S}_{111}^{(j)}(2,t)
 & = &
 - \frac{6 \mu^2}{j!} \ln^j\left( \frac{\left(t-m^2\right)\left(t-9m^2\right)}{3 m^3 \sqrt{t}} \right)
 +
 \frac{\left(t+3m^2\right)^4}{4 t \left(t-m^2\right)\left(t-9m^2\right)} \tilde{S}_{111}^{(j-2)}(2,t),
 \nonumber \\
\eq
with the convention that $\tilde{S}_{111}^{(j)}(2,t)=0$ for $j<0$.

The simple form of the differential equation in eq.~(\ref{dgl_tilde_2_2eps}) and in eq.~(\ref{dgl_transformed}) for $\tilde{S}_{111}$ is the key to the iterative solution.
However, we would like to mention that there is a small price to pay:
The original function $S_{111}(2-2\eps,t)$ is regular at $t=0$.
This is no longer the case for $\tilde{S}_{111}(2-2\eps,t)$, which exhibits logarithmic singularities at $t=0$.
This is due to the fact that in the definition of $\tilde{S}_{111}$ we split off a prefactor
\bq
 e^{\frac{\eps}{2} \ln t}
 & = &
 1 + \frac{\eps}{2} \ln t + {\mathcal O}\left(\eps^2\right).
\eq
Of course, in the combination of prefactor and $\tilde{S}_{111}$ all logarithmic singularities of $\ln(t)$ cancel,
leaving a regular result.

\section{Boundary values}
\label{sec:boundary_values}

In order to obtain the two-loop sunrise integral from the differential equation we need the boundary values 
at some point $t=t_0$. It is advantageous to choose for $t_0$ a value where the elliptic curve is degenerated, so that
the sunrise integral at $t_0$ can expressed in terms of ordinary multiple polylogarithms.
A possible choice is $t_0=0$.
We have
\bq
 S_{111}\left( 2-2\eps, 0\right)
 & = &
 \Gamma\left(1+2\eps\right)
 \left(\frac{m^2}{\mu^2}\right)^{-1-2\eps}
 \int\limits_{\sigma} \frac{\omega}{\left( x_1 + x_2 + x_3 \right)^{1+2\eps} {\cal U}^{1-\eps}}.
\eq
By a change of variables we can relate this integral 
to the one-loop three-point function 
in $4+2\eps$ space-time dimensions (please note the sign of the $\eps$-part)
with massless internal lines
and three external masses.
The change of variables can be found in \cite{Adams:2013nia} and the result of the one-loop three point function can be taken from
\cite{Bern:1994kr,Lu:1992ny}.
One obtains for the $\eps$-expansion of $S_{111}( 2-2\eps, 0)$
\bq
 \sum\limits_{j=0}^\infty \eps^j S_{111}^{(j)}\left(2,0\right)
 & = &
 e^{2 \gamma \eps} \Gamma\left(1+2\eps\right)
 \left( \frac{m^2\sqrt{3}}{\mu^2} \right)^{-1-2\eps}
 \left[ \frac{3}{2\eps^2} \frac{\Gamma\left(1+\eps\right)^2}{\Gamma\left(1+2\eps\right)}
        f
        - \frac{\pi}{\eps} \right],
\eq
with
\bq
 f
 & = &
 \frac{1}{i}
 \left[ 
  \left(-r_3\right)^{-\eps} \; {}_2F_1\left(-2\eps,-\eps;1-\eps; r_3 \right)
  -
  \left(-r_3^{-1}\right)^{-\eps} \; {}_2F_1\left(-2\eps,-\eps;1-\eps; r_3^{-1} \right)
 \right].
\eq
The hypergeometric function can be expanded systematically in $\eps$ with the methods of \cite{Moch:2001zr}.
The first few terms are given by
\bq
 {}_2F_1\left(-2\eps,-\eps;1-\eps; x \right)
 & = &
 1 + 2 \eps^2 \mathrm{Li}_2\left(x\right)
 + \eps^3 \left[ 2 \mathrm{Li}_3\left(x\right) - 4 \mathrm{Li}_{2,1}\left(x,1\right) \right]
 \nonumber \\
 & &
 + \eps^4 \left[ 2 \mathrm{Li}_4\left(x\right) - 4 \mathrm{Li}_{3,1}\left(x,1\right) + 8 \mathrm{Li}_{2,1,1}\left(x,1,1\right) \right]
 + {\mathcal O}\left(\eps^5\right).
\eq
We obtain for the first few terms of the Taylor expansion
\bq
\lefteqn{
 S_{111}^{(0)}\left(2,0\right)
 = 
 \frac{\sqrt{3} \mu^2}{i m^2}
 \left[ \mathrm{Li}_2\left(r_3\right) - \mathrm{Li}_2\left(r_3^{-1}\right) \right],
} & &
 \\
\lefteqn{
 S_{111}^{(1)}\left(2,0\right)
 = 
 \frac{\sqrt{3} \mu^2}{i m^2}
 \left\{
  - 2 \mathrm{Li}_{2,1}\left(r_3,1\right) - \mathrm{Li}_3\left(r_3\right) 
  + 2  \mathrm{Li}_{2,1}\left(r_3^{-1},1\right) + \mathrm{Li}_3\left(r_3^{-1}\right)
 \right\}
 } & & 
 \nonumber \\
 & &
  - 2 \ln\left(\frac{m^2 \sqrt{3}}{\mu^2}\right) S_{111}^{(0)}\left(2,0\right),
 \nonumber \\
\lefteqn{
 S_{111}^{(2)}\left(2,0\right)
 = 
 \frac{\sqrt{3} \mu^2}{i m^2}
 \left\{
    4 \mathrm{Li}_{2,1,1}\left(r_3,1,1\right) - 2 \mathrm{Li}_{3,1}\left(r_3,1\right) + \mathrm{Li}_4\left(r_3\right) 
  - 4 \mathrm{Li}_{2,1,1}\left(r_3^{-1},1,1\right) 
 \right. } & & \nonumber \\
 & &
 \left.
  + 2 \mathrm{Li}_{3,1}\left(r_3^{-1},1\right) 
  - \mathrm{Li}_4\left(r_3^{-1}\right) 
  + \frac{2\pi^2}{9} \left[
                           \mathrm{Li}_2\left(r_3\right) - \mathrm{Li}_2\left(r_3^{-1}\right) 
                    \right]
 \right\}
  - 2 \ln\left(\frac{m^2 \sqrt{3}}{\mu^2}\right) 
        S_{111}^{(1)}\left(2,0\right)
 \nonumber \\
 & & 
  - 2 \ln^2\left(\frac{m^2 \sqrt{3}}{\mu^2}\right) S_{111}^{(0)}\left(2,0\right).
 \nonumber
\eq
We may determine the boundary values for $\tilde{S}_{111}(2-2\eps,t)$ as well. As already mentioned, the
function $\tilde{S}_{111}$ has logarithmic singularities at $t=0$, which are removed by the prefactor
relating $\tilde{S}_{111}$ and $S_{111}$. It will be convenient to express these logarithms in $\ln(-q)$.
We have
\bq
 \lim\limits_{t\rightarrow 0}
 \left(
 e^{\frac{\eps}{2}\ln\left(-q\right)}
 \sum\limits_{j=0}^\infty \eps^j \tilde{S}_{111}^{(j)}\left(2,t\right)
 \right)
 & = &
 e^{2\eps\ln\left(\frac{m^2}{\mu^2}\right)-2\sum\limits_{n=2}^\infty \frac{\left(-1\right)^n}{n} \zeta_n \eps^n}
 \sum\limits_{j=0}^\infty \eps^j S_{111}^{(j)}\left(2,0\right).
\eq
Explicitly, we have for the lowest terms the asymptotic expansions
\bq
 \tilde{S}_{111}^{(0)}\left(2,0\right) 
 & = &
 S_{111}^{(0)}\left(2,0\right),
 \nonumber \\
 \tilde{S}_{111}^{(1)}\left(2,t\right) 
 & \sim &
 S_{111}^{(1)}\left(2,0\right)
 + \left[ - \frac{1}{2} \ln\left(-q\right) + 2 \ln\left(\frac{m^2}{\mu^2}\right) \right] S_{111}^{(0)}\left(2,0\right),
 \nonumber \\
 \tilde{S}_{111}^{(2)}\left(2,t\right) 
 & \sim &
 S_{111}^{(2)}\left(2,0\right)
 + \left[ - \frac{1}{2} \ln\left(-q\right) + 2 \ln\left(\frac{m^2}{\mu^2}\right) \right] S_{111}^{(1)}\left(2,0\right)
 \nonumber \\
 & &
 + \left[ \frac{1}{8} \ln^2\left(-q\right) - \ln\left(\frac{m^2}{\mu^2}\right) \ln\left(-q\right) 
          + 2 \ln^2\left(\frac{m^2}{\mu^2}\right) - \zeta_2 \right] S_{111}^{(0)}\left(2,0\right).
\eq

\section{The class of functions}
\label{sec:functions}

In this section we present the class of functions needed to express our results.
In particular we are interested in the $q$-dependence of these functions.
Let us start from known functions and let us consider first the $q$-independent functions.
Here we have the algebraic functions extended by the multiple logarithms.
We recall that the classical polylogarithms are defined by
\bq
 \mathrm{Li}_n\left(x\right) & = & \sum\limits_{j=1}^\infty \; \frac{x^j}{j^n},
\eq
and that the multiple polylogarithms are defined by
\bq
\label{def_multiple_polylogs}
 \mathrm{Li}_{n_1,n_2,...,n_k}\left(x_1,x_2,...,x_k\right)
 & = &
 \sum\limits_{j_1=1}^\infty \sum\limits_{j_2=1}^{j_1-1} ... \sum\limits_{j_k=1}^{j_{k-1}-1}
 \frac{x_1^{j_1}}{j_1^{n_1}} \frac{x_2^{j_2}}{j_2^{n_2}} ... \frac{x_k^{j_k}}{j_k^{n_k}}.
\eq
Let us now turn to the $q$-dependent functions:
In previous publications \cite{Adams:2014vja,Adams:2015gva} we already introduced 
the following generalisation of the classical polylogarithm 
depending on three variables $x$, $y$, $q$ and two (integer) indices $n$, $m$:
\bq
 \mathrm{ELi}_{n;m}\left(x;y;q\right) & = & 
 \sum\limits_{j=1}^\infty \sum\limits_{k=1}^\infty \; \frac{x^j}{j^n} \frac{y^k}{k^m} q^{j k}.
\eq
For the results of this paper we just need one more generalisation, given by
\bq
\lefteqn{
 \mathrm{ELi}_{n_1,...,n_l;m_1,...,m_l;2o_1,...,2o_{l-1}}\left(x_1,...,x_l;y_1,...,y_l;q\right) 
 = }
 & & \nonumber \\
 & = &
 \sum\limits_{j_1=1}^\infty ... \sum\limits_{j_l=1}^\infty
 \sum\limits_{k_1=1}^\infty ... \sum\limits_{k_l=1}^\infty
 \;\;
 \frac{x_1^{j_1}}{j_1^{n_1}} ... \frac{x_l^{j_l}}{j_l^{n_l}}
 \;\;
 \frac{y_1^{k_1}}{k_1^{m_1}} ... \frac{y_l^{k_l}}{k_l^{m_l}}
 \;\;
 \frac{q^{j_1 k_1 + ... + j_l k_l}}{\prod\limits_{i=1}^{l-1} \left(j_i k_i + ... + j_l k_l \right)^{o_i}}.
\eq
In addition we will use the shorthand notation
\bq
 \mathrm{ELi}_{\vec{n};\vec{m};\vec{o}}\left(\vec{x};\vec{y};q\right)
 & = & 
 \mathrm{ELi}_{n_1,...,n_l;m_1,...,m_l;2o_1,...,2o_{l-1}}\left(x_1,...,x_l;y_1,...,y_l;q\right),
\eq
where the vectors $\vec{x}$, $\vec{y}$, $\vec{n}$ and $\vec{m}$ have $l$ entries, while the vector
$\vec{o}$ has $(l-1)$ entries.
In this paper the entries of $\vec{o}$ will always be even numbers and it is therefore convenient to use the convention
$\vec{o}=(2o_1,...,2o_{l-1})$.
There is no need to include a $l$-th entry in the vector $\vec{o}$, the effect of including $2o_l$
can be re-absorbed by $n_l'=n_l+o_l$ and $m_l'=m_l+o_l$.

\section{The integration algorithm}
\label{sec:algorithm}

Let us consider the differential equation
\bq
\label{basic_dgl}
 L^{(0)}_{2} F(t)
 & = &
 \mu^2 I(t),
\eq
where the differential operator $L^{(0)}_2$ is defined in eq.~(\ref{def_L2_expansion}).
Let us fix the boundary conditions at the value $t_0$.
The full solution can be written as
\bq
\label{inhomogeneous_solution}
 F\left(t\right)
 & = & 
 C_1\left(t_0\right) \psi_1\left(t\right) + C_2\left(t_0\right) \psi_2\left(t\right)
 + F_{\mathrm{special}}\left(t,t_0\right).
\eq
The special solution $F_{\mathrm{special}}\left(t,t_0\right)$ satisfies
\bq
 F_{\mathrm{special}}\left(t_0,t_0\right) & = & 0.
\eq
The integration constants are determined from the boundary conditions.
The special solution $F_{\mathrm{special}}\left(t,t_0\right)$ is given by
\bq
 F_{\mathrm{special}}\left(t,t_0\right)
 & = &
  \mu^2 \int\limits_{t_0}^{t} dt_1 \frac{I(t_1)}{p_2(t_1) W(t_1)} \left[ - \psi_1(t) \psi_2(t_1) + \psi_2(t) \psi_1(t_1) \right].
\eq
There are two alternative representations for $F_{\mathrm{special}}(t)$, which will be useful \cite{Adams:2014vja}.
We have
\bq
 F_{\mathrm{special}}\left(t,t_0\right)
 & = &
 -
 \frac{\psi_1}{\pi}
 \int\limits_{q_0}^q \frac{dq_1}{q_1}
 \int\limits_{q_0}^{q_1} \frac{dq_2}{q_2}
 \frac{\mu^2 \psi_1(q_2)^3}{\pi p_2(q_2) W(q_2)^2}
 I(q_2)
 \nonumber \\
 & = &
 -
 \frac{\psi_1}{\pi}
 \frac{1}{2i}
 \sum\limits_{j=1}^\infty
 \sum\limits_{k=1}^\infty
 \int\limits_{q_0}^q \frac{dq_1}{q_1}
 \int\limits_{q_0}^{q_1} \frac{dq_2}{q_2}
 \left(r_3^j - r_3^{-j} \right) 
 k^2 \left(-1\right)^k
 \left(-q_2\right)^{jk}
 I(q_2).
\eq
We would like to fix the boundary values at $t=0$. However, since $\tilde{S}_{111}$ has logarithmic singularities
at $t=0$, this is not directly possible.
We first consider the boundary values at a small, but finite value $t=t_0$ (or equivalently $q=q_0$) and take
the limit $t_0 \rightarrow 0$ (or equivalently $q_0 \rightarrow 0$) in the end.
Before taking the limit we are allowed to neglect any polynomials in $t_0$ (or $q_0$)
and we only need to keep logarithms of $t_0$ (respectively logarithms of $q_0$).

Let us further mention a small technical detail:
The homogeneous solutions are spanned by $\psi_1$ and $\psi_2$, defined in eq.~(\ref{def_periods}).
For the case at hand it will be convenient to use instead of the basis $\{\psi_1,\psi_2\}$
the basis given by
\bq
 \psi_1,
 & &
 \psi_1 \ln\left(-q\right)
\eq
in eq.~(\ref{inhomogeneous_solution}).

Now let us specialise to $\tilde{S}_{111}^{(j)}(2,t)$.
We have
\bq
 L^{(0)}_{2} \tilde{S}_{111}^{(j)}(2,t)
 & = &
 - \frac{6 \mu^2}{j!} \ln^j\left( \frac{\left(t-m^2\right)\left(t-9m^2\right)}{3 m^3 \sqrt{t}} \right)
 +
 \frac{\left(t+3m^2\right)^4}{4 t \left(t-m^2\right)\left(t-9m^2\right)} \tilde{S}_{111}^{(j-2)}(2,t),
 \nonumber \\
\eq
with the convention that $\tilde{S}_{111}^{(j)}(2,t)=0$ for $j<0$.
The inhomogeneous term splits into two parts
\bq
 I(t) & = &
 I^{(j)}_a(t) + I^{(j)}_b(t),
\eq
with
\bq
\label{def_I_a_I_b}
 I^{(j)}_a(t) 
 & = & 
 - \frac{6}{j!} \ln^j\left( \frac{\left(t-m^2\right)\left(t-9m^2\right)}{3 m^3 \sqrt{t}} \right),
 \nonumber \\
 I^{(j)}_b(t)
 & = & 
 \frac{\left(t+3m^2\right)^4}{4 \mu^2 t \left(t-m^2\right)\left(t-9m^2\right)} \tilde{S}_{111}^{(j-2)}(2,t).
\eq
At each order $j$ the inhomogeneous term $I^{(j)}_a(t)$ gives a new contribution, while
the inhomogeneous term $I^{(j)}_b(t)$ yields an iterated integration of an already existing term of lower order.

In the next two sub-sections we will show that the integrations of $I^{(j)}_a(t)$ and $I^{(j)}_b(t)$
can be done within the class of functions defined in section~\ref{sec:functions}.

\subsection{Integration of $I^{(j)}_a(t)$}

We first consider the integration of $I^{(j)}_a(t)$:
\bq
\tilde{S}_{111}^{(j,a)}
 & = &
 -
 \frac{\psi_1}{\pi}
 \frac{1}{2i}
 \sum\limits_{j_1=1}^\infty
 \sum\limits_{k_1=1}^\infty
 \int\limits_{q_0}^q \frac{dq_1}{q_1}
 \int\limits_{q_0}^{q_1} \frac{dq_2}{q_2}
 \left(r_3^{j_1} - r_3^{-j_1} \right) 
 k_1^2 \left(-1\right)^{k_1}
 \left(-q_2\right)^{j_1 k_1}
 I^{(j)}_a(t),
\eq
where $I^{(j)}_a(t)$ is given in eq.~(\ref{def_I_a_I_b}).
The $q$-expansion of the logarithm is given by
\bq
\label{q_expansion_logarithm}
\lefteqn{
 \ln\left( \frac{\left(t-m^2\right)\left(t-9m^2\right)}{3 m^3 \sqrt{t}} \right)
 =
 -\frac{1}{2} \ln\left(-q\right)
 + 12 \mathrm{ELi}_{1;0}\left(-1;1;-q\right)
 } \\
 & &
 + \mathrm{ELi}_{1;0}\left(r_3;-1;-q\right)
 + \mathrm{ELi}_{1;0}\left(r_3^{-1};-1;-q\right)
 -3 \mathrm{ELi}_{1;0}\left(r_3;1;-q\right)
 -3 \mathrm{ELi}_{1;0}\left(r_3^{-1};1;-q\right).
 \nonumber
\eq
Taking the $j$-th power of eq.~(\ref{q_expansion_logarithm}), we see 
that the integration is of the form
\bq
\lefteqn{
 \int\limits_{q_0}^q \frac{dq_1}{q_1}
 \int\limits_{q_0}^{q_1} \frac{dq_2}{q_2}
 \ln^k\left(-q_2\right)
 \mathrm{ELi}_{0;-2}\left(x_0;y_0;-q_2\right)
  \prod\limits_{l=1}^{j-k}
    \mathrm{ELi}_{1;0}\left(x_l;y_l;-q_2\right)
 = } & &
 \nonumber \\
 & &
 \int\limits_{-q_0}^{-q} \frac{dq_1}{q_1}
 \int\limits_{-q_0}^{q_1} \frac{dq_2}{q_2}
 \ln^k\left(q_2\right)
 \mathrm{ELi}_{0;-2}\left(x_0;y_0;q_2\right)
  \prod\limits_{l=1}^{j-k}
    \mathrm{ELi}_{1;0}\left(x_l;y_l;q_2\right).
\eq
The basic integral is (with $n \in {\mathbb N}$, $k \in {\mathbb N}_0$)
\bq
 \int\limits \frac{dq}{q}
 q^n \ln^k\left(q\right)
 & = &
 \sum\limits_{r=0}^k
 \left(-1\right)^r
 \frac{k!}{(k-r)!}
 \frac{q^n}{n^{r+1}} \ln^{k-r}\left(q\right).
\eq
Each logarithm is always accompanied by a positive power of $q$.
This allows us to take the limit $q_0\rightarrow 0$ without any problems.
We obtain for $k<j$
\bq
\label{int_I_a_1}
\lefteqn{
 \int\limits_{0}^q \frac{dq_1}{q_1}
 \int\limits_{0}^{q_1} \frac{dq_2}{q_2}
 \ln^k\left(-q_2\right)
 \mathrm{ELi}_{0;-2}\left(x_0;y_0;-q_2\right)
  \prod\limits_{l=1}^{j-k}
    \mathrm{ELi}_{1;0}\left(x_l;y_l;-q_2\right)
 = } & &
 \nonumber \\
 & = &
 \sum\limits_{r=0}^k 
 \left(-1\right)^r
 \frac{\left(r+1\right)k!}{\left(k-r\right)!}
 \ln^{k-r}\left(-q\right)
 \nonumber \\
 & &
 \times
 \mathrm{ELi}_{0,1,...,1;-2,0,...,0;4+2r,0,...,0}\left(x_0,x_1,...,x_{j-k};y_0,y_1,...,y_{j-k};-q\right).
 \eq
In the special case $k=j$ we simply have
\bq
\label{int_I_a_2}
\lefteqn{
 \int\limits_{0}^q \frac{dq_1}{q_1}
 \int\limits_{0}^{q_1} \frac{dq_2}{q_2}
 \ln^k\left(-q_2\right)
 \mathrm{ELi}_{0;-2}\left(x_0;y_0;-q_2\right)
 = } & &
 \nonumber \\
 & = &
 \sum\limits_{r=0}^k 
 \left(-1\right)^r
 \frac{\left(r+1\right)k!}{\left(k-r\right)!}
 \ln^{k-r}\left(-q\right)
 \mathrm{ELi}_{2+r;r}\left(x_0;y_0;-q\right).
 \eq
Eq.~(\ref{int_I_a_1}) and eq.~(\ref{int_I_a_2}) show
that the integrations related to $I^{(j)}_a(t)$ always stay within our class of functions.

\subsection{Integration of $I^{(j)}_b(t)$}

Let us now consider the integration of $I^{(j)}_b(t)$:
\bq
\tilde{S}_{111}^{(j,b)}
 & = &
 -
 \frac{\psi_1}{\pi}
 \int\limits_{q_0}^q \frac{dq_1}{q_1}
 \int\limits_{q_0}^{q_1} \frac{dq_2}{q_2}
 \frac{\mu^2 \psi_1(q_2)^3}{\pi p_2(q_2) W(q_2)^2}
 I^{(j)}_b(q_2),
\eq
where $I^{(j)}_b(q)$ is given in eq.~(\ref{def_I_a_I_b}) and involves the lower-order terms $\tilde{S}_{111}^{(j-2)}\left(2,t\right)$.
The function $\tilde{S}_{111}^{(j-2)}\left(2,t\right)$ contains always a prefactor $\psi_1/\pi$ and it is convenient to write
\bq
 \tilde{S}_{111}^{(j-2)}\left(2,t\right)
 & = & 
 \frac{\psi_1}{\pi}
 \tilde{E}_{111}^{(j-2)}\left(2,q\right).
\eq
Let us define the integration kernel 
\bq
 H(q) & = &
 - \frac{\mu^2 \psi_1^3}{\pi p_2 W^2}
 \frac{\left(t+3m^2\right)^4}{4 \mu^2 t \left(t-m^2\right)\left(t-9m^2\right)}
 \frac{\psi_1}{\pi}.
\eq
$\tilde{S}_{111}^{(j,b)}$ is then given by
\bq
 \tilde{S}_{111}^{(j,b)}
 & = &
 \frac{\psi_1}{\pi}
 \int\limits_{q_0}^q \frac{dq_1}{q_1}
 \int\limits_{q_0}^{q_1} \frac{dq_2}{q_2}
 H\left(q_2\right)
 \tilde{E}_{111}^{(j-2)}\left(2,q_2\right).
\eq
The integration kernel has the $q$-expansion
\bq
 H 
 & = &
 \frac{1}{4}
 \left\{ 
  1 - 2 \sqrt{3} i \left[ 
                          \mathrm{ELi}_{0;0}\left(r_3;1;-q\right) 
                          -
                          \mathrm{ELi}_{0;0}\left(r_3^{-1};1;-q\right) 
                   \right]
 \right\}^4.
\eq
The terms in $\tilde{E}_{111}^{(j-2)}\left(2,q\right)$ are of the form (with $k \in {\mathbb N}_0$ and $l \in {\mathbb N}_0$)
\bq
\label{Etilde_general_form}
 \ln^k\left(-q\right) \mathrm{ELi}_{n_1,n_2,...,n_l;m_1,m_2,...,m_l;2o_1,2o_2,...,2o_{l-1}}\left(x_1,x_2,...,x_l;y_1,y_2,...,y_l;-q\right),
\eq
including the case $l=0$, in which case eq.~(\ref{Etilde_general_form}) reduces to
\bq
 \ln^k\left(-q\right).
\eq
The integration we have to do is given by
\bq
 I & = &
 \int\limits_{q_0}^q \frac{dq_1}{q_1}
 \int\limits_{q_0}^{q_1} \frac{dq_2}{q_2}
 \ln^k\left(-q_2\right) 
 \left(
  \prod\limits_{t=1}^{p}
    \mathrm{ELi}_{0;0}\left(x_t;y_t;-q_2\right)
 \right)
 \nonumber \\
 & &
 \times
 \mathrm{ELi}_{n_{p+1},...,n_{p+l};m_{p+1},...,m_{p+l};2o_{p+1},...,2o_{p+l-1}}\left(x_{p+1},...,x_{p+l};y_{p+1},...,y_{p+l};-q_2\right),
\eq
with $p \in \{0,1,2,3,4\}$.
Let us first consider the case $(p,l) \neq (0,0)$. 
In this case 
each logarithm is accompanied by a positive power of $q$ and we may take the 
limit $q_0\rightarrow 0$ without any problems.
We obtain
\bq
 I 
 & = &
 \sum\limits_{r=0}^k 
 \left(-1\right)^r
 \frac{\left(r+1\right)k!}{\left(k-r\right)!}
 \ln^{k-r}\left(-q\right)
 \mathrm{ELi}_{\vec{n};\vec{m};\vec{o}}\left(x_{1},...,x_{p+l};y_{1},...,y_{p+l};-q\right),
\eq
where
\bq
 \vec{n} & = & ( \underbrace{0,0,...,0}_{p},n_{p+1},...,n_{p+l} ),
 \nonumber \\
 \vec{m} & = & ( \underbrace{0,0,...,0}_{p},m_{p+1},...,m_{p+l} ),
 \nonumber \\
 \vec{o} & = & ( 4+2r,\underbrace{0,...,0}_{p-1},2o_{p+1},...,2o_{p+l-1} ).
\eq
This leaves the case $(p,l)=(0,0)$. In this case we have to keep the $q_0$-dependence.
The integral is rather simple and we have
\bq
\label{log_terms}
 \int\limits_{q_0}^q \frac{dq_1}{q_1}
 \int\limits_{q_0}^{q_1} \frac{dq_2}{q_2}
 \ln^k\left(-q_2\right) 
 & = &
 \frac{1}{\left(k+1\right)\left(k+2\right)} \left[ \ln^{k+2}\left(-q\right) - \ln^{k+2}\left(-q_0\right) \right]
 \nonumber \\
 & &
 - \frac{1}{\left(k+1\right)} \ln^{k+1}\left(-q_0\right) \left[ \ln\left(-q\right) - \ln\left(-q_0\right) \right].
\eq
Again we see that all integrations can be carried out within our class of functions.

At this point a few comments on the $\ln(-q)$-terms and $\ln(-q_0)$-terms are in order:
The $\ln(q_0)$-terms from eq.~(\ref{log_terms}) will cancel with the corresponding $\ln(q_0)$-terms
from the integration constants $C_1$ and $C_2$ appearing in eq.~(\ref{inhomogeneous_solution}).
The $\ln(-q)$-terms remain in the result for $\tilde{S}_{111}^{(j)}$. 
They are removed by the prefactor of eq.~(\ref{def_Stilde}), once we convert to $S_{111}^{(j)}$.

\section{Conclusions}
\label{sec:conclusions}

In this paper we have shown, that the Laurent expansion around $D=2-2\eps$ space-time dimensions
of the two-loop sunrise integral with equal non-zero masses can be computed to arbitrary order
in the dimensional regularisation parameter $\eps$.
We have defined a class of transcendental functions and we have shown that all results 
can be expressed within this class of functions.
The class of functions includes the multiple polylogarithms, which are functions associated to
a punctured Riemann surface of genus zero.
The new additional functions are associated to
an algebraic variety consisting of a punctured Riemann surface of genus one and a punctured Riemann surface of genus zero.
We provided an algorithm which allows us to express an arbitrary order of the $\eps$-expansion of the two-loop sunrise integral with
equal non-zero masses in terms of these functions.
We expect these functions to be useful for other Feynman integrals as well.

\subsection*{Acknowledgements}

L.A. is grateful for financial support from the research training group GRK 1581.
C.B. thanks Deutsche Forschungsgemeinschaft for financial support under the project BO4500/1-1 and Humboldt University for hospitality.


\begin{appendix}


\section{Explicit results}
\label{sec:results}

In this appendix we give the explicit results for the first three terms of the $\eps$-expansion of $S_{111}(2,t)$.
In order to write the results in a compact form we introduce specific linear combinations 
of the functions defined in section~\ref{sec:functions}.
We define a prefactor $c_n$ and a sign $s_n$, both depending on an index $n$ by
\bq
 c_n = \frac{1}{2} \left[ \left(1+i\right) + \left(1-i\right)\left(-1\right)^n\right] = 
 \left\{ \begin{array}{rl}
 1, & \mbox{$n$ even}, \\
 i, & \mbox{$n$ odd}, \\
 \end{array} \right.
 & &
 s_n = (-1)^n =
 \left\{ \begin{array}{rl}
 1, & \mbox{$n$ even}, \\
 -1, & \mbox{$n$ odd}. \\
 \end{array} \right.
\eq
At depth $1$ we define the linear combinations
\bq
\lefteqn{
 \mathrm{E}_{n;m}\left(x;y;q\right) 
 = } & & 
 \\
 & = &
 \frac{c_{n+m}}{i}
 \left[
  \left( \frac{1}{2} \mathrm{Li}_n\left( x \right) + \mathrm{ELi}_{n;m}\left(x;y;q\right) \right)
  - s_{n+m}
  \left( \frac{1}{2} \mathrm{Li}_n\left( x^{-1} \right) + \mathrm{ELi}_{n;m}\left(x^{-1};y^{-1};q\right)
  \right)
 \right].
 \nonumber
\eq
More explicitly, we have
\bq
\label{def_classical_E}
\lefteqn{
 \mathrm{E}_{n;m}\left(x;y;q\right) 
 = } & & \\
 & = &
 \left\{ \begin{array}{ll}
 \frac{1}{i}
 \left[
 \frac{1}{2} \mathrm{Li}_n\left( x \right) - \frac{1}{2} \mathrm{Li}_n\left( x^{-1} \right)
 + \mathrm{ELi}_{n;m}\left(x;y;q\right) - \mathrm{ELi}_{n;m}\left(x^{-1};y^{-1};q\right)
 \right],
 & \mbox{$n+m$ even,} \\
 & \\
 \frac{1}{2} \mathrm{Li}_n\left( x \right) + \frac{1}{2} \mathrm{Li}_n\left( x^{-1} \right)
 + \mathrm{ELi}_{n;m}\left(x;y;q\right) + \mathrm{ELi}_{n;m}\left(x^{-1};y^{-1};q\right),
 & \mbox{$n+m$ odd.} \\
 \end{array}
 \right.
 \nonumber
\eq
The functions $\mathrm{E}_{n;m}(x;y;q)$ can be thought of as elliptic generalisations of the Clausen and Glaisher functions \cite{Adams:2015gva}.
At higher depth we define functions
\bq
 \mathrm{E}_{n_1,...,n_l;m_1,...,m_l;2o_1,...,2o_{l-1}}\left(x_1,...,x_l;y_1,...,y_l;q\right) 
\eq
as follows:
If $o_1=0$ we set
\bq
\lefteqn{
 \mathrm{E}_{n_1,...,n_l;m_1,...,m_l;0,2o_2,...,2o_{l-1}}\left(x_1,...,x_l;y_1,...,y_l;q\right) 
 = } & & \nonumber \\
 & &
 \mathrm{E}_{n_1;m_1}\left(x_1;y_1;q\right) 
 \mathrm{E}_{n_2,...,n_l;m_2,...,m_l;2o_2,...,2o_{l-1}}\left(x_2,...,x_l;y_2,...,y_l;q\right).
\eq
For $o_1 \neq 0$ we set
\bq
\label{E_higher_depth_integral_repr}
\lefteqn{
 \mathrm{E}_{n_1,...,n_l;m_1,...,m_l;2o_1,...,2o_{l-1}}\left(x_1,...,x_l;y_1,...,y_l;q\right) 
 = 
 \int\limits_0^q \frac{dq_1}{q_1} \int\limits_0^{q_1} \frac{dq_2}{q_2} ... \int\limits_0^{q_{o_1-1}} \frac{dq_{o_1}}{q_{o_1}}
 } & & \\
 & &
 \left[ \mathrm{E}_{n_1; m_1}\left(x_1;y_1;q_{o_1}\right) - \mathrm{E}_{n_1; m_1}\left(x_1;y_1;0\right) \right]
 \mathrm{E}_{n_2,...,n_l;m_2,...,m_l;2o_2,...,2o_{l-1}}\left(x_2,...,x_l;y_2,...,y_l;q_{o_1}\right).
 \nonumber
\eq
The integrals are easily converted to sums. 
For example we have
\bq
\lefteqn{
 \mathrm{E}_{0,1;-2,0;4}\left(x_1,x_2;y_1,y_2;-q\right)
 = 
 } & & \\
 & = & 
 \int\limits_0^{q} \frac{dq_1}{q_1} \int\limits_0^{q_1} \frac{dq_2}{q_2} 
 \left[ \mathrm{E}_{0; -2}\left(x_1;y_1;-q_2\right) - \mathrm{E}_{0; -2}\left(x_1;y_1;0\right) \right]
 \mathrm{E}_{1; 0}\left(x_2;y_2;-q_2\right)
 \nonumber \\
 & = &
 \frac{1}{i}
 \left\{
 \left[ \mathrm{ELi}_{2; 0}\left(x_1;y_1;-q\right) - \mathrm{ELi}_{2; 0}\left(x_1^{-1};y_1^{-1};-q\right) \right]
 \times
 \frac{1}{2} 
 \left[ \mathrm{Li}_{1}\left(x_2\right) + \mathrm{Li}_{1}\left(x_2^{-1}\right) \right]
 \right. \nonumber \\
 & & \left.
 + \mathrm{ELi}_{0,1;-2,0;4}\left(x_1,x_2;y_1,y_2;-q\right)
 - \mathrm{ELi}_{0,1;-2,0;4}\left(x_1^{-1},x_2;y_1^{-1},y_2;-q\right)
 \right. \nonumber \\
 & & \left.
 + \mathrm{ELi}_{0,1;-2,0;4}\left(x_1,x_2^{-1};y_1,y_2^{-1};-q\right)
 - \mathrm{ELi}_{0,1;-2,0;4}\left(x_1^{-1},x_2^{-1};y_1^{-1},y_2^{-1};-q\right)
 \right\}.
 \nonumber
\eq
Certain terms in the result at order $\eps^j$ will be proportional to lower order terms.
The factor of proportionality is
\bq
 L_{1;0}
 & = &
         - 2 \ln\left( \frac{m^2}{\mu^2} \right) 
         - \mathrm{E}_{1;0}\left(r_3;-1;-q\right) 
         + 3 \mathrm{E}_{1;0}\left(r_3;1;-q\right) 
         - 6 \mathrm{E}_{1,0}\left(-1;1;-q\right).
\eq
It is convenient to factor out the homogeneous solution and 
we write
\bq
 S_{111}^{(j)}\left(2,t\right)
 & = & 
 \frac{\psi_1}{\pi}
 E_{111}^{(j)}\left(2,q\right).
\eq
We now present the explicit results for the functions $E_{111}^{(j)}(2,q)$ for $j \in \{0,1,2\}$.
We have
\bq
\lefteqn{
 E^{(0)}_{111}
 = 
 3 \mathrm{E}_{2;0}\left(r_3;-1;-q\right),
 } & &
 \nonumber \\
\lefteqn{
 E^{(1)}_{111}
 = 
 3 \mathrm{E}_{3;1}\left(r_3;-1;-q\right)
 + 3 \mathrm{E}_{0,1;-2,0;4}\left(r_3,r_3;-1,-1;-q\right)
 - 9 \mathrm{E}_{0,1;-2,0;4}\left(r_3,r_3;-1,1;-q\right)
 } & &
 \nonumber \\
 & &
 + 18 \mathrm{E}_{0,1;-2,0;4}\left(r_3,-1;-1,1;-q\right)
 + \frac{3}{2i} \left\{ 
                       - 2 \mathrm{Li}_{2,1}\left(r_3,1\right) - 2 \mathrm{Li}_3\left(r_3\right)
                       + 2 \mathrm{Li}_{2,1}\left(r_3^{-1},1\right) 
 \right. \nonumber \\
 & & \left.
                       + 2 \mathrm{Li}_3\left(r_3^{-1}\right)
                       + 6 \mathrm{Li}_1\left(-1\right) \left[ \mathrm{Li}_2\left(r_3\right) - \mathrm{Li}_2\left(r_3^{-1}\right) \right]
                \right\}
 + L_{1;0} E^{(0)}_{111},
 \nonumber \\
\lefteqn{
 E^{(2)}_{111}
 =
 \frac{9}{4} \mathrm{E}_{4;2}\left(r_3;-1;-q\right)
 + 108 \mathrm{E}_{2,0,0,0,0;0,0,0,0,0;4,0,0,0}\left(r_3,r_3,r_3,r_3,r_3;-1,1,1,1,1;-q\right)
 } & &
 \nonumber \\
 & &
 + 108 \mathrm{SE}_{0,0,0,0;0,0,0,0;4,0,0}\left(r_3,r_3,r_3,r_3;1,1,1,1;-q\right)
   \frac{1}{2i} \left[ \mathrm{Li}_2\left(r_3\right) - \mathrm{Li}_2\left(r_3^{-1}\right) \right]
 \nonumber \\
 & &
 + 3 \mathrm{E}_{0,1;-2,0;6}\left(r_3,r_3;-1,-1;-q\right)
 - 9 \mathrm{E}_{0,1;-2,0;6}\left(r_3,r_3;-1,1;-q\right)
 \nonumber \\
 & &
 +18 \mathrm{E}_{0,1;-2,0;6}\left(r_3,-1;-1,1;-q\right)
 \nonumber \\
 & &
 + \frac{27}{2} \mathrm{E}_{0,1,1;-2,0,0;4,0}\left(r_3,r_3,r_3;-1,1,1;-q\right)
 - 9            \mathrm{E}_{0,1,1;-2,0,0;4,0}\left(r_3,r_3,r_3;-1,-1,1;-q\right)
 \nonumber \\
 & &
 + \frac{3}{2}  \mathrm{E}_{0,1,1;-2,0,0;4,0}\left(r_3,r_3,r_3;-1,-1,-1;-q\right)
 - 54           \mathrm{E}_{0,1,1;-2,0,0;4,0}\left(r_3,r_3,-1;-1,1,1;-q\right)
 \nonumber \\
 & &
 + 18           \mathrm{E}_{0,1,1;-2,0,0;4,0}\left(r_3,r_3,-1;-1,-1,1;-q\right)
 + 54           \mathrm{E}_{0,1,1;-2,0,0;4,0}\left(r_3,-1,-1;-1,1,1;-q\right)
 \nonumber \\
 & &
 + \frac{3}{2i} \left\{
                        4 \mathrm{Li}_{2,1,1}\left(r_3,1,1\right) - 2 \mathrm{Li}_{3,1}\left(r_3,1\right) + \frac{1}{4} \mathrm{Li}_4\left(r_3\right)
                        -4 \mathrm{Li}_{2,1,1}\left(r_3^{-1},1,1\right) + 2 \mathrm{Li}_{3,1}\left(r_3^{-1},1\right) 
 \right. \nonumber \\
 & & \left.
                        - \frac{1}{4} \mathrm{Li}_4\left(r_3^{-1}\right)
                        + 6 \mathrm{Li}_1\left(-1\right) \left[ 
                                                               - 2 \mathrm{Li}_{2,1}\left(r_3,1\right) - \mathrm{Li}_3\left(r_3\right)
                                                               + 2 \mathrm{Li}_{2,1}\left(r_3^{-1},1\right) + \mathrm{Li}_3\left(r_3^{-1}\right)
                                                         \right]
 \right. \nonumber \\
 & & \left.
                        + 18 \left(\mathrm{Li}_1\left(-1\right)\right)^2 \left[ 
                                                                                \mathrm{Li}_2\left(r_3\right)
                                                                              - \mathrm{Li}_2\left(r_3^{-1}\right)
                                                                         \right]
                \right\}
 + \zeta_2 
   \frac{1}{2i} \left[ \mathrm{Li}_2\left(r_3\right) - \mathrm{Li}_2\left(r_3^{-1}\right) \right]
 \nonumber \\
 & &
 + L_{1;0} E^{(1)}_{111}
 - \frac{1}{2} \left( L_{1;0} \right)^2 E^{(0)}_{111}
 + \zeta_2 E^{(0)}_{111}.
\eq
In the expression for $E^{(2)}_{111}$ we used in addition the abbreviation
\bq
\lefteqn{
 \mathrm{SE}_{0,0,0,0;0,0,0,0;4,0,0}\left(r_3,r_3,r_3,r_3;1,1,1,1;-q\right)
 = 
 \mathrm{E}_{0,0,0,0;0,0,0,0;4,0,0}\left(r_3,r_3,r_3,r_3;1,1,1,1;-q\right)
} & & \nonumber \\
 & &
 + \frac{1}{2i} \left[ \mathrm{Li}_0\left(r_3\right) - \mathrm{Li}_0\left(r_3^{-1}\right)  \right]
                \mathrm{E}_{0,0,0;0,0,0;4,0}\left(r_3,r_3,r_3;1,1,1;-q\right)
 \nonumber \\
 & &
 - \frac{1}{4} \left[ \mathrm{Li}_0\left(r_3\right) - \mathrm{Li}_0\left(r_3^{-1}\right)  \right]^2
                \mathrm{E}_{0,0;0,0;4}\left(r_3,r_3;1,1;-q\right)
 \nonumber \\
 & &
 - \frac{1}{8i} \left[ \mathrm{Li}_0\left(r_3\right) - \mathrm{Li}_0\left(r_3^{-1}\right)  \right]^3
                \frac{1}{i} \left[ \mathrm{ELi}_{2;2}\left(r_3;1;-q\right) - \mathrm{ELi}_{2;2}\left(r_3^{-1};1;-q\right) \right].
 \hspace*{30mm}
\eq
The function $\mathrm{SE}$ is symmetric in the four four-tuples $(n_i,m_i,x_i,y_i)=(0,0,r_3,1)$.

We have verified analytically the correctness of these results by re-inserting the results into the original
differential equation.
In addition we verified numerically the results by comparing with the program {\verb|sector_decomposition|} \cite{Bogner:2007cr}.

\end{appendix}

\bibliography{/home/stefanw/notes/biblio}
\bibliographystyle{/home/stefanw/latex-style/h-physrev5}

\end{document}